\documentclass{iaus}
\usepackage{graphicx}

\title[RIAF  in  LLAGN]  
{Radiatively   inefficient   accretion   disks   in   Low   Luminosity
AGN\thanks{Based  on  observations  obtained  at the  Space  Telescope
Science   Institute,  which   is  operated   by  the   Association  of
Universities  for  Research  in  Astronomy, Incorporated,  under  NASA
contract NAS 5-26555.}}

\author[Macchetto \& Chiaberge]   
{F. D. Macchetto$^1$, M. Chiaberge$^1$}%

\affiliation{$^1$Space Telescope Science Institute, 3700 San Martin Dr., Baltimore, MD 21218 \break email: macchetto@stsci.edu}

\pubyear{2006}
\volume{238}  
\pagerange{001--999}
\date{??? and in revised form ???}
\jname{Black Holes: from Stars to Galaxies -- across the Range of Masses}
\editors{V. Karas \& G. Matt, eds.}
\begin{document}

\maketitle

\begin{abstract}
We study  a complete and distance-limited  sample of 25  LINERs, 21 of
which have been imaged with  the Hubble Space Telescope to study their
physical properties and to  compare their radio and optical properties
with those of other samples of local AGNs, namely Seyfert galaxies and
low-luminosity  radio  galaxies (LLRG).   Our  results  show that  the
LINERs population is not homogeneous,  as there are two subclasses: i)
the first  class is similar to  LLRG, as it extends  the population of
radio-loud nuclei to lower luminosities;  ii) the second is similar to
Seyferts, and extends the properties of radio-quiet nuclei towards the
lowest luminosities.  The different nature  of the various  classes of
local  AGN are  best understood  when  the fraction  of the  Eddington
luminosity they irradiate $L_o/L_{Edd}$ is plotted against the nuclear
radio-loudness parameter: Seyferts are associated with relatively {\it
high} radiative  efficiencies $L_o/L_{Edd} \gtrsim  10^{-4}$ (and high
accretion rates onto {\it low}  mass black holes); LLRG are associated
with {\it  low} radiative efficiencies  (and low accretion  rates onto
{\it  high}  black  hole   masses);  all  LINERs  have  low  radiative
efficiency (and accretion rates), and can be radio-loud or radio quiet
depending on their black hole mass.


\end{abstract}

\firstsection 

\section{Introduction}
Low  luminosity active  galactic  nuclei (LLAGN)  are  believed to  be
powered  by accretion of  matter onto  the central  supermassive black
hole, similarly  to powerful  AGN. In a  large fraction of  LLAGN, the
central  black hole  is  as  massive as  in  powerful distant  quasars
($M_{BH}  \sim 10^{8}- 10^{9}  M_\odot$), thus  their very  low nuclear
luminosity  implies  that accretion  occurs  with  very low  radiative
efficiency     (or     at      very     low      rates) \cite{ho04coz,papllagn}. If so, the physics of the accretion process
may be different from  the ``standard'' optically thick, geometrically
thin  accretion disks. Because of the very
low radiation they emit at  all wavelengths, these objects are  very difficult  to observe. While the  AGN nature of optical nuclear components seen in HST images of a sample of
LLAGN have been  unambiguously established it  is still unclear whether the  radiation is from a
jet or from the accretion flow. LLAGN have also been found to lie
on  the  so-called  ``fundamental   plane  of  black  hole  activity''
\cite{merloni03,falcke04}, which attempts  to unify the emission from
all  sources around  black holes,  over a  large range  of  masses and
luminosities,  from  Galactic sources  to  powerful  quasars. But  the
origin of such  a ``fundamental plane'' and its  relationship with the
origin   of   the   radiation    is   still   a   matter   of   debate.

\section{LINERs in the framework of the local AGN population}

Different  accretion disk  models  are expected  to  show the  largest
difference  in  spectral  shape in the IR-to-UV region.   RIAFs  should lack  both  the  "big
blue-bump" and  the IR (reprocessed) bump,  which instead characterize
optically thick,  geometrically thin  accretion disk emission  and the
surrounding  heated  dust.   For  example, in  low  luminosity  radio
galaxies  non-thermal  emission from  the  jet  dominates the  optical
nuclear  radiation,  \cite{pap1}, while  the  Galactic  center is  not
visible in the optical because it is hidden by a large amount of dust.

We have studied  a complete and distance-limited sample  of 25 LINERs,
21 of which have been imaged with the Hubble Space Telescope.  In nine
objects  we detect  an unresolved  nucleus.  In  order to  study their
physical properties,  we compare the  radio and optical  properties of
the nuclei of LINERs with those of other samples of local AGNs, namely
Seyfert  galaxies  and   low-luminosity  FR~I  radio  galaxies  (LLRG)
(Fig.~1).  The radio-optical correlation found for FR~I, which is best
explained as the result of a  single emission process in the two bands
(i.e.   non-thermal   synchrotron  emission  from  the   base  of  the
relativistic jet), provides us with a powerful tool to investigate the
origin of the  nuclei.  We have shown that  in the radio-optical plane
of the nuclei  there is a clear separation  between Seyferts and radio
galaxies.    For  similar   radio  core   luminosity,   Seyfert~1  are
significantly brighter in the  optical than FR~I.  Therefore, although
most Seyferts have $R=L_{5GHz}/L_B>10$, radio-quiet and radio-loud AGN
appear to be still well  differentiated. This implies that the nuclear
physical properties  of the  two classes are  significantly different.
Our results  show that  the LINERs population  is not  homogeneous, as
there are two subclasses: i) the first class is similar to LLRG, as it
extends the population of radio-loud nuclei to lower luminosities; ii)
the  second is  similar to  Seyferts,  and extends  the properties  of
radio-quiet nuclei towards  the lowest luminosities.  Furthermore, all
radio-loud  LINERs have $M_{BH}/M\odot  \gtrsim 10^8$,  while Seyferts
and radio-quiet LINERs have $M_{BH}/M\odot \lesssim 10^8$.

\begin{figure}
 \begin{minipage}[b]{0.5\linewidth}
\includegraphics[height=2.0in,width=2.0in,angle=0]{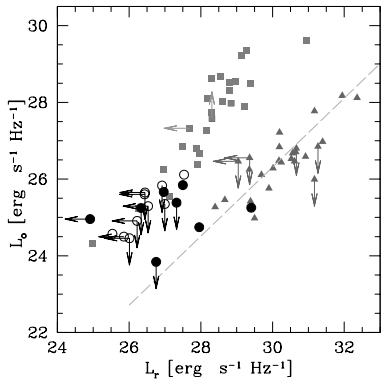}
 \end{minipage}
\begin{minipage}[b]{0.5\linewidth}
\caption{Optical nuclear luminosity vs. radio core luminosity for 
LINERs   (circles), Seyferts    (squares)  and  FR~I  radio   galaxies
(triangles).  The   dashed  line is the    correlation between the two
quantities found  for 3CR  FR~I  sample. Open  circles  are LINERs  in
late-type hosts, filled circles are LINERs in early-type hosts.}
 \end{minipage}
\end{figure}

We  have derived  the radiative  efficiency of  the  accretion process
around the  central black holes in  our samples of local  AGN.  All of
them emit only a small fraction of the Eddington luminosity.  Although
the determination of the bolometric luminosity is uncertain because of
the lack  of detailed spectral  information, the accretion  process in
LINERs  appears  to  take  place  on  a  highly  sub-Eddington  regime
($L_o/L_{Edd} < 10^{-5}$, and can  be as low as $\sim 10^{-8}$).  Such
low values  are clearly  not compatible with  the expectations  from a
standard   optically  thick   and  geometrically   thin  (quasar-like)
accretion  disk.   Thus, low  accretion  rates  and/or low  efficiency
processes  appears to  be required  in  all LINERs.   Our results  are
qualitatively  in agreement  with \cite{ho04coz},  who made  use  of the
$H\alpha$ emission line  as an indicator of the AGN  power for a large
sample of Seyferts and LINERs.

The  different  nature of the various  classes  of local  AGN are best
understood  when  the    fraction of the    Eddington luminosity  they
irradiate is plotted against the  nuclear radio-loudness (Fig.~2). 
Our objects  populate  three   different  quadrants,
according   to their physical  properties.  We   identify Seyferts and
radio-quiet LINERs as  the high and  low accretion  rate counterparts,
respectively.  For   low accretion regimes,  the  nuclei appears to be
``radio-loud'' only when  a more massive  black hole ($M_{BH} > 10^{8}
M_\odot$)  is  present. We speculate that  the  fourth quadrant, which
appears    to be  ``empty''  in   the   local universe, would  contain
radio-loud  nuclei  with high $L_o/L_{Edd}$,  readily  identified with
radio loud quasars.

We  we  further  tested  this  picture  by studying  the nuclear spectral energy distribution
of a galaxy,  NGC~4565, that seems to be  a perfect candidate
for hosting  a RIAF around  the central supermassive black  hole.  The
object is part of the ``Palomar sample'' of LLAGN \cite{ho97}, and it
is included in both the \cite{merloni03} and \cite{falcke04} samples
that  were  used to  define  the  ``fundamental  plane of  black  hole
activity''. It  is worth  mentioning that NGC~4565  does not  show any
significant peculiarity  in that plane.  NGC~4565 is  a nearby (d=9.7
Mpc)  LLAGN  classified  as  a  Seyfert~1.9 because  of  the  possible
presence  of a  faint,  relatively  broad (FWHM  =  1750 km  s$^{-1}$)
H$\alpha$ line. Although it  is  a Type  2  Seyfert, this  object is  only moderately  absorbed, and  the  nuclear radiation  is  visible in  the
optical spectral region.  NGC~4565  may thus represent the first clear
example of low-luminosity accretion  onto a supermassive black hole in
the optical band.

\begin{figure}
 \begin{minipage}[b]{0.5\linewidth}
\includegraphics[height=2.1in,width=2.1in,angle=0]{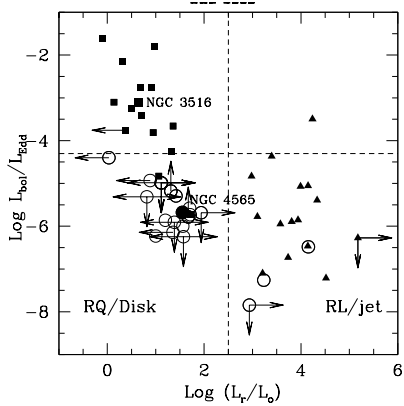}
 \end{minipage}
 \begin{minipage}[b]{0.5\linewidth}
\caption{Optical  to Eddington  luminosity ratio  plotted  against the
radio to optical ratio (the ``nuclear radio-loudness'') for the sample
of  nearby  LLAGN  (\cite{papllagn}).   Seyfert~1s  are
plotted  as  squares, low  luminosity  radio  galaxies are  triangles,
LINERs  are  empty circles.   NGC~4565  (filled  circle) and  the Seyfert~1 galaxy NGC~3516
(large square)  are also marked in  the figure.  The  dashed lines are
only  used  to guide  the  eye  and divide  objects  of  high and  low
Eddington  ratio  (top  and  bottom  of the  figure)  and  radio-quiet
(disk-dominated  nuclei, left)  or radio  loud (jet  dominated nuclei,
right). \label{r_edd}}
 \end{minipage}
\end{figure}

The SED (Fig.~3) is peculiar, as it is almost flat in a $\log \nu -
log (\nu  F\nu)$ representation, with  no sign of  both a UV  bump and
thermally  reprocessed IR emission.   The very  low luminosity  of the
source associated with a relatively high central black hole mass imply
an  extremely small value  of the  Eddington ratio  ($L_o/L_{Edd} \sim
10^{-6}$). This, together with the position occupied by this object on
diagnostic planes  for low  luminosity AGN, represents  clear evidence
for  a low  radiative  efficiency  accretion process  at  work in  its
innermost  regions.   

\begin{figure}
 \begin{minipage}[b]{0.5\linewidth}
\includegraphics[height=2.3in,width=2.3in,angle=0]{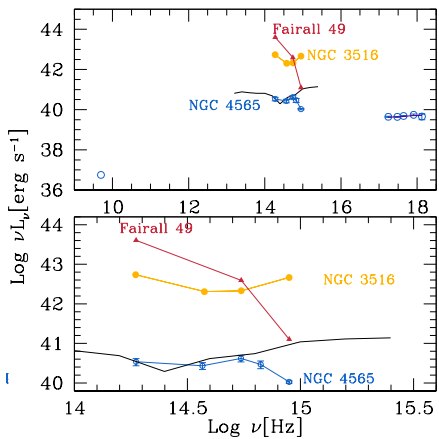}
 \end{minipage}
\hspace{1.0cm}
 \begin{minipage}[b]{0.4\linewidth}
\caption{Absorption corrected nuclear  spectral energy distribution of
NGC~4565 from  the radio  to the X-ray  band.  The X-ray  spectrum has
been  significantly re-binned to  improve the  clarity of  the figure.
The solid  line superimposed to the  X-ray data is a spectral model
used to fit the data.  For comparison, we show the IR-to-UV
SED  of  a  Seyfert~1  (NGC~3516)  and  of  a  Compton-thin  Seyfert~2
(Fairall~49). The  solid line  is the average  SED of  radio-quiet QSO
from \cite{elvis94}, normalized to the  flux of NGC~4565 in the F814W
filter.   The  lower  panel  is  a zoom  into  the  IR-to-UV  spectral
region. \label{sed}}
 \end{minipage}
\end{figure}

The fact  that the  [OIII] emission line  flux is substantial  in this
object implies that  an extended narrow line region,  similar to other
Seyfert galaxies, is still  present in NGC~4565. A possible intriguing
scenario  is  that the  active  nucleus  has recently  ``turned-off'',
switching  from a  high  efficiency, standard,  accretion  disk, to  a
radiative inefficient accretion process.  However, since the EW of the
[OIII]5007 emission  line is  rather small, with  the present  data we
cannot rule out  that the amount of ionizing photons  from the RIAF is
sufficient to produce the observed [OIII] flux.
\section{Conclusions}

The scenario we propose  needs  further investigation, since   optical
detections of the nuclei are  available for a  minority of the LINER's
sample.  Thus, deep imaging  with high spatial  resolution (achievable
only with the Hubble Space Telescope) are crucial. In particular, when
suitable observations of  a large number  of LINERs will be available,
it will be possible to address whether  the dichotomy persists or some
of the low black hole mass objects with optical  upper limits mix with
the population of ``radio-loud''  LINERs.  Clearly this would  falsify
our scenario  for the role  of the black hole  mass in determining the
radio-loudness  of  the nuclei.   Deep  imaging  of a  larger complete
sample would  also address  the issue  of whether there  is continuous
transition between the two classes. This  is indeed a subject of great
interest in  the  study of high  luminosity  quasars  and for   a more
complete understanding of the  overall subject should be also extended
to lower end of the AGN luminosity function.

\begin{acknowledgments}
We acknowledge R.Gilli, A. Capetti and W. Sparks for a
productive collaboration.
\end{acknowledgments}

\end{document}